\documentclass[12pt]{article}
\pdfoutput=1
\usepackage{comment}
\usepackage{amsmath}
\usepackage{amssymb}
\usepackage{nccmath}
\usepackage{graphicx}
\usepackage{here}
\usepackage{subcaption}
\usepackage{url}
\usepackage[sort&compress, numbers, merge]{natbib}
\usepackage{braket}
\usepackage{physics}
\usepackage[compat=1.1.0]{tikz-feynhand}
\usepackage{slashed}
\usepackage{mathtools}
\usepackage{bbold}

\usepackage{todonotes}

\newcommand{\1}{\mathbb{1}}
\newcommand{\alias}[3]{%
  \let#3=#2
  \newcommand{#1}{#3}
  \renewcommand{#2}{{\textcolor{red}{#3}}}
}
\alias{\unw}{\beta}{\origbeta}
\newcommand{\cdA}{D}
\newcommand{\cdF}{D}

\newcommand{\defby}{\coloneqq}
\newcommand{\lagrangian}{\mathcal{L}}
\newcommand{\hamiltonian}{\mathcal{H}}
\newcommand{\Vlambda}{\tilde{\lambda}}
\newcommand{\Hlambda}{\lambda}
\newcommand{\Seucl}{S_\mathrm{Eucl}}
\alias{\ct}{\dagger}{\origdagger}
\newcommand{\fund}{h}

\newcommand{\delrho}[1][]{\partial_\rho^{#1}}
\newcommand{\rhoE}{\rho_{\mathrm{E}}}
\newcommand{\delrhoE}[1][]{\partial_{\rhoE}^{#1}}
\newcommand{\delunw}[1][]{\partial_\unw^{#1}}
\newcommand{\tE}{t_{\mathrm{E}}}
\newcommand{\eff}[1]{\mathcal{#1}_{\mathrm{eff}}}
\newcommand{\Keff}{\eff{K}}
\newcommand{\Veff}{T}
\newcommand{\meff}{m_\text{eff}}
\newcommand{\const}{\text{const.}}
\newcommand{\fgamma}{f^{\gamma}_\unw}
\newcommand{\fW}{f^W_\unw}
\newcommand{\adjprofile}{\Phi}
\newcommand{\brad}{{\rhoE^*}}
\newcommand{\U}{\mathrm{U}}
\newcommand{\SU}{\mathrm{SU}}
\newcommand{\pauli}[1]{\tau^{#1}}
\newcommand{\sugen}[1]{\frac{\pauli{#1}}{2}}
\newcommand{\adjmass}{m_{\phi}}
\newcommand{\numrange}[2]{{#1}\text{--}{#2}}

\numberwithin{equation}{section}

\newcommand{\version}{arXiv2}
\DeclareRobustCommand{\modifiedat}[2]{%
\ifthenelse{\equal{\version}{#1}}{\textcolor{red}{#2}}{#2}%
}

\usepackage[colorlinks=true, linkcolor=blue, citecolor=blue,
urlcolor=black]{hyperref} 
\usepackage[capitalize]{cleveref}
\crefname{section}{Sec.}{Secs.}
                 
\begin{document}
\def\ps{\mathbf{p}}
\def\PS{\mathbf{P}}
\baselineskip 0.6cm
\def\simgt{\mathrel{\lower2.5pt\vbox{\lineskip=0pt\baselineskip=0pt
           \hbox{$>$}\hbox{$\sim$}}}}
\def\simlt{\mathrel{\lower2.5pt\vbox{\lineskip=0pt\baselineskip=0pt
           \hbox{$<$}\hbox{$\sim$}}}}
\def\simprop{\mathrel{\lower3.0pt\vbox{\lineskip=1.0pt\baselineskip=0pt
             \hbox{$\propto$}\hbox{$\sim$}}}}
\def\tr{\mathop{\rm tr}}
\def\SU{\mathop{\rm SU}}

\def\VG{V}
\def\VH{v}
\def\azimuth{\varphi}
\newcommand{\az}{\azimuth}
\def\zenith{\theta}
\def\higgsprofile{\xi}
\def\Newton{G_N}

\begin{titlepage}

\begin{flushright}
IPMU23-0054
\end{flushright}

\vskip 1.1cm

\begin{center}

{\Large \bf 
Revisiting Metastable Cosmic String Breaking
}

\vskip 1.2cm
Akifumi Chitose $^{a}$, 
Masahiro Ibe$^{a,b}$,
Yuhei Nakayama$^{a}$, 
Satoshi Shirai$^{b}$, and
Keiichi Watanabe$^{a}$ 
\vskip 0.5cm

{\it

$^a$ {ICRR, The University of Tokyo, Kashiwa, Chiba 277-8582, Japan}

$^b$ {Kavli Institute for the Physics and Mathematics of the Universe
 (WPI), \\The University of Tokyo Institutes for Advanced Study, \\ The
 University of Tokyo, Kashiwa 277-8583, Japan}

}

\vskip 1.0cm

\abstract{
Metastable cosmic strings appear in models of new physics with a two-step symmetry breaking $G\to H\to 1$, where
$\pi_1(H)\neq 0$
and $\pi_1(G)=0$.
They decay via the monopole-antimonopole pair creation inside.
Conventionally, the breaking rate has been estimated by an infinitely thin string approximation, which requires a large hierarchy between the symmetry breaking scales.
In this paper, we reexamine it by taking into account the finite sizes of both the cosmic string and the monopole.
We obtain a robust lower limit on the tunneling factor $e^{-S_B}$ even for regimes the conventional estimate is unreliable.
In particular, it is relevant to the cosmic string interpretation of the gravitational wave signals recently reported by pulsar timing array experiments.
}

\end{center}
\end{titlepage}

\section{Introduction} 
Topological defects arising from cosmological phase transitions have been studied intensely (see e.g., Ref.~\cite{Vilenkin:2000jqa}).
A nontrivial fundamental group of a vacuum manifold leads to linear defects called cosmic strings.
In particular, a spontaneous breaking of a $\U$(1) symmetry leads to a string solution, since $\pi_1$($\U$(1))$= \mathbb{Z}$.
Their evolution and consequences in the early Universe are indispensable topics in the new physics search, as various extensions of the Standard Model have $\U(1)$ symmetry.

One of the most promising potential observation channels is the gravitational waves (GWs)
(for reviews see, e.g., Refs.~\cite{Vilenkin:1982hm,Hindmarsh:2011qj,Auclair:2019wcv,Gouttenoire:2019kij}).
While topological, infinitely long strings are stable, loops and small scale structures can release energy primarily as gravitational waves.
Also, repeated mutual interactions of strings lead to so-called scaling regime of the string network, in which a certain number of cosmic strings remain in the Hubble horizon.
Thus, gravitational waves are continuously emitted and may be observed as the stochastic gravitational background.

\modifiedat{JHEP2}{Recently, multiple pulsar timing array (PTA) collaborations} reported such stochastic GW 
signal exhibiting the Hellings–Downs angular correlation in the nHz range~\modifiedat{JHEP2}{\cite{NANOGrav:2023gor,EPTA:2023fyk,Reardon:2023gzh,Xu:2023wog}}.
Intriguingly, the observed spectrum 
favors metastable cosmic strings
over stable ones as its origin~\cite{NANOGrav:2023hvm}.
The metastability of the string results in the suppression of the low-frequency spectrum of the GW in the PTA band (see also Ref.\,\cite{Leblond:2009fq,Buchmuller:2019gfy,Buchmuller:2020lbh,Buchmuller:2021mbb,Buchmuller:2023aus} for theoretical
 works on the GW spectrum from metastable strings).
The observations by PTAs ignited tremendous research interests in metastable cosmic strings~\cite{Lazarides:2023ksx,Fu:2023mdu,Lazarides:2023rqf,Maji:2023fhv,Afzal:2023kqs,Servant:2023tua,Buchmuller:2023aus}.

Metastable strings appear in models with
successive symmetry breaking, e.g.,
\begin{align}
    G \stackrel{\VG}{\rightarrow} \U(1) \stackrel{\VH}{\rightarrow} \mbox{nothing} \ ,
\end{align}
with hierarchical 
vacuum expectation values (VEVs), $\VG\gg \VH$~\cite{Preskill:1992ck}.
Here, we presume $\pi_1(G) = 0$
while $\pi_2(G/\U(1))=\pi_1(\U(1))= \mathbb{Z}$.
In this class of models,
cosmic string solutions appear as classically stable configurations in the low-energy 
effective theory, while they turn out to be metastable in the full theory.
In fact, these models possess
a monopole configuration associated with the first stage of the symmetry breaking. 
The strings can be cut via Schwinger production of a monopole-antimonopole pair inside, which is a tunneling process.

Conventionally,
the string breaking process is approximated by a bubble formation of the monopole worldline on the string worldsheet, where the size and the thickness of the monopole and the string are taken to be zero.
In particular, neglecting the monopole radius corresponds to assuming large hierarchy.
In this semiclassical approximation,
the resultant breaking rate per string unit length is given in Refs.\,\cite{Vilenkin:1982hm,Preskill:1992ck} as
\begin{align}
    \Gamma &= \frac{\mu}{2\pi} e^{-\pi \kappa} \ , \quad \kappa  = \frac{m_M^2}{\mu} \ ,
\end{align}
where $\mu$ and $m_M$ are the string tension and the monopole mass, respectively.
By using this formula,
the PTA data
can be well fitted for $\sqrt{\kappa} \simeq 8$ and $ -7\lesssim \log_{10}\Newton\mu\lesssim -4$~\cite{NANOGrav:2023hvm}, where $\Newton$ is the Newton constant.

In this paper, we revisit 
the estimation of the string breaking rate in light of PTAs' GW observations.
The GW spectrum in the PTA band from the metastable strings crucially depends on the breaking rate.
Although the conventional approximation assumes large hierarchy between the breaking scales,
it is not very significant for 
$\sqrt{\kappa} \sim 8$
favored by the observation.
Thus, the validity of the approximation
is quite unclear.
To remedy this situation, we reanalyze the string breaking rate using the Ans\"atze on the string unwinding process proposed by Ref.\,\cite{Shifman:2002yi}.
An effective two dimensional field theory on the string world sheet is constructed with
the soliton sizes taken into account, thereby revealing their influence on the breaking rate.

As we will see, for a large hierarchy ($\sqrt{\kappa}\gg 1$), our estimate of the bounce action $S_B$, which gives the tunneling factor as $e^{-S_B}$, agrees with the conventional one up to a factor of $O(1)$.
We also obtain a robust lower limit on the tunneling factor $e^{-S_B}$ even for regimes the conventional estimate is unreliable.

The rest of this paper is organized as follows.
In \cref{sec:SetUp}, 
we state the setup of our analysis.
In \cref{sec:Ansatz},
we introduce the unwinding Ansatz
proposed in Ref.\,\cite{Shifman:2002yi}.
In \cref{sec:Numerical}, we perform numerical estimate of the bounce action.
The final section is devoted to the discussions and conclusions.

\section{Model with Hierarchical Breaking}
\label{sec:SetUp}
\subsection{Setup}
For simplicity,
we limit ourselves to an SU(2) gauge theory with
an adjoint scalar field $\phi^a$ ($a=1,2,3$)
and a doublet scalar 
field $h_i$ ($i=1,2$).
Following Ref.\,\cite{Shifman:2002yi}, 
we take the Lagrangian as
\begin{equation} \label{eq:lagrangian}
\lagrangian=-\frac{1}{2}(\cdA_\mu \phi^a)(\cdA^\mu \phi^a) - (\cdF_\mu \fund)^\ct (\cdF^\mu \fund) - \frac{1}{4g^2}F^a_{\mu\nu}F^{a\mu\nu}\ -V_\mathrm{Higgs}(\phi,\fund)\ ,
\end{equation}
where $g$ is the gauge coupling constant. We take the Minkowski metric as $(g_{\mu\nu})=(-1,1,1,1)$.
The covariant derivatives are given by
\begin{align}
    \cdF_\mu \fund&\defby\partial_\mu \fund - i \sugen{a}A_\mu^a\fund\ ,\\
    \cdA_\mu \phi^a &\defby \partial_\mu\phi^a + \epsilon^{abc}A_\mu^b \phi^c\ ,
\end{align}
where $\pauli{}$'s are the Pauli matrices and the doublet indices are suppressed.
The scalar potential is given by
\begin{equation}
    V_{\mathrm{Higgs}}(\phi,\fund)\defby \Hlambda\pqty{\abs{\fund}^2-v^2}^2+\Vlambda\pqty{\phi^a\phi^a-V^2}^2+\gamma \abs{\pqty{\phi-\frac{V}{2}}\fund}^2\ ,
\end{equation}
where $\phi\defby\phi^a\pauli{a}/2$.
The dimensionless coupling constants $\Hlambda$, $\Vlambda$ 
and $\gamma$ are taken to be positive.
We also assume that the two mass scales $\VG$  and $\VH$ are hierarchical, i.e., $\VG \gg \VH$.

At the vacuum, $\phi$ takes the trivial configuration
\begin{equation}
\label{eq:VEV1}
   \langle{\phi^a}\rangle = \VG \delta^{a3}
\end{equation}
without loss of generality.
This breaks SU(2) down to U(1).
The remaining U(1) corresponds to the SO(2) rotation about the $a=3$ axis of SU(2).

Due to  $\gamma>0$, the second component of $h$ obtains a large mass squared, and hence, the VEV of $h$ is given by
\begin{align}
    \langle h\rangle =  
    \mqty(\VH\\0)\ ,
\end{align}
which breaks the remaining U(1) symmetry.
As the U(1) charge of $h$ is $\pm 1/2$,
this breaks $\mathbb{Z}_2$ center symmetry of SU(2).

\subsection{Monopoles}
At the first phase transition SU(2)$\to$U(1),
the 't Hooft-Polyakov monopole appears as a topological defect~\cite{tHooft:1974kcl,Polyakov:1974ek}. 
The static monopole configuration
with a unit winding number at the origin is given by
\begin{align}
\label{eq:heg}
\phi^a=\VG H(r)\displaystyle{\frac{x^a}{r}}\ , \quad
A_{0}^a = 0 \ ,\quad
A^{a}_{i}=
\displaystyle{\frac{\epsilon^{aij}x^j}{r^2}}F(r)\ ,\quad (i,j=1,2,3)\ ,
\end{align}
where $(x^0,x^1,x^2,x^3)$
are the spacetime coordinates and 
$r\defby\sqrt{x_1^2+x_2^2+x_3^2}$.
Assuming the hierarchy between
$v$ and $V$, we neglect the effect of $h$, and set it to zero.
The profile functions $H(r)$ and $F(r)$ satisfy the boundary conditions
\begin{alignat}{2}
H(r)&\to 0 ,\, (r\to0)\ , &\qquad H(r)&\to 1\ ,\, (r\to \infty)\ ,\\
F(r)&\to 0 , \, (r\to 0)\ , & \qquad F(r)&\to 1\ ,\, (r\to \infty)\ ,
\end{alignat}
where they approach their limits exponentially at $r\to \infty$.

To see the magnetic field,
it is convenient to define the effective U(1)
field strength as
\begin{equation}
\label{eq:effectiveF}
F^{\mathrm{U}(1)}_{\mu\nu} := \frac{1}{\VG}\phi^a F^{a}_{\mu\nu}
\end{equation}
(see e.g., Ref.~\cite{Shifman:2012zz}).
The only non-vanishing components of $F^{\U(1)}{}^{\mu\nu}$ are 
\begin{equation}
    F^{\U(1)}{}^{i j } = -
\frac{\epsilon^{ijk}x^k}{r^3} (2F-F^2)H \ , \quad (i,j=1,2,3)\ .
\end{equation}
Hence, the magnetic charge of the monopole is given by
\begin{equation}
\label{eq: monopole flux}
    Q^m:=  \int_{r \to \infty} 
    \dd
S_{i} B^{\U(1)}{}^{i  } =  -4\pi \ ,
\end{equation}
where $B^{\U(1)}{}^i =\varepsilon^{ijk}F^{\U(1)}_{jk}/2$ and $\dd{S}_{i}$ is the surface element of a two dimensional sphere surrounding the monopole.

\begin{figure}[tb]
\centering
  \includegraphics[width=0.7\linewidth]{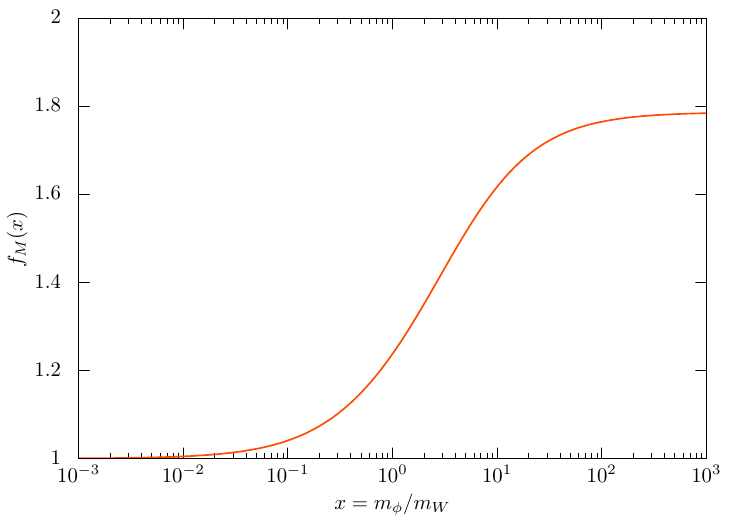}
\caption{Monopole mass function $f_M(x)$ with $x = \adjmass/m_W$.}
\label{fig:monopole_mass}
\end{figure}

The monopole mass $m_M$ may be parameterized as ~\cite{Bogomolny:1976ab,Kirkman:1981ck},
\begin{align}
    m_M = \frac{4\pi \VG}{g} f_M\qty(\frac{\adjmass}{m_W})\ ,
\end{align}
where $\adjmass$ and $m_W$ are the adjoint Higgs mass and the SU(2) gauge boson mass, respectively:
\begin{align}
    \adjmass = \sqrt{8\Vlambda} \VG \ , \quad m_W = g \VG \ . 
\end{align}
\Cref{fig:monopole_mass} shows $f_M(x)$ calculated numerically.
It can be well approximated by
\begin{align}
    f_M(x)=\frac{1 + 16.8264 x + 33.7119 x^2+12.0448 x^3}{1+16.3264 x+27.3047 x^2+6.74157 x^3}\ .
\end{align}
The limiting values are $f_M(x\to 0) = 1$ and $f_M(x\to \infty)\simeq 1.787$.

\subsection{Cosmic Strings}
\label{sec:CosmicString}
Let us now discuss the 
Abrikosov-Nielsen-Olesen (ANO)
string formed at the second symmetry breaking
assuming the vacuum of $\phi$ in Eq.\,\eqref{eq:VEV1}.
In the large $\VG$ limit, $A^{1,2}$, $\phi$ and $h_2$ decouple from $A^3$ and $h_1$.
Hence, we may treat the low-energy effective theory as a U(1) gauge theory with a complex scalar field $h_1$ with charge $1/2$. 
The covariant derivative is given by
\begin{align}
    D_\mu h_1 = \qty(\partial_\mu -  \frac{i}{2}A_\mu^3)h_1 \ .
\end{align}

The static string solution along the $x^3$-axis has the form (see e.g., Ref.~\cite{Vilenkin:2000jqa})
\begin{align}
\label{eq:string ansatz1}
h(\rho) &= \mqty(h_1(\rho)\\0)\  ,\\ 
h_1(\rho) &=\VH\higgsprofile(\rho)e^{-i n\azimuth}\ , \\
\label{eq:string ansatz2}
A_{i}^3&={2n}\frac{\epsilon_{ij}x^j}{\rho^2}(1-f(\rho))\ ,~~~~(i,j=1,2)\ , \\
A^3_{0} &= A^3_{3} = 0 \ ,
\end{align}
where $n \in \mathbb{Z}$ is the winding number of the string and $\higgsprofile(\rho)$ and $f(\rho)$ are the profile functions.
Here, the second component $\fund_2$
obtains a large mass, $m_{\fund_2} \defby \sqrt{\gamma} \VG$, and thus we take $\fund_2 = 0$.
We adopted the cylindrical coordinates $\azimuth\defby{\arctan}(x_2/x_1)$ and $\rho\defby\sqrt{x_1^2+x_2^2}$.
The two-dimensional antisymmetric tensor is defined so that $\epsilon_{12}=1$.%
    \footnote{Noting that $\dd{\azimuth} = - \dd{x}^i \epsilon_{ij}x^j/\rho^2$, 
Eq.\,\eqref{eq:string ansatz2} can be rewritten as $A_{i} \dd{x}^i = n \times f(\rho)\dd{\azimuth}$.} 
The boundary conditions for the profile functions are
\begin{align}
\higgsprofile(\rho)\rightarrow 0\ , ~(\rho\rightarrow0)&\ ,~~~~~\higgsprofile(\rho)\rightarrow 1\ , ~(\rho\rightarrow \infty)\ ,\\
f(\rho)\rightarrow 1\ , ~(\rho\rightarrow 0)&\ ,~~~~~f(\rho)\rightarrow 0\ , 
~(\rho\rightarrow \infty)\ .
\end{align}
They approach unity for $\rho \gg (g\VH)^{-1}$ exponentially.
Also, $D_\mu h_1$ approaches zero exponentially,
making the string tension finite.

The winding number is related to the magnetic flux along the string by
\begin{align}
\label{eq: string flux}
\int \dd[2]{x}B_{3}=\oint_{\rho\rightarrow \infty} A_{i}^3 \dd{x}^i=-{4\pi n}\ .
\end{align}
Thus, for $|n|=1$, the magnetic flux along a string coincides with the magnetic charge of a magnetic (anti)monopole.

\begin{figure}[tb]
\centering
  \includegraphics[width=0.7\linewidth]{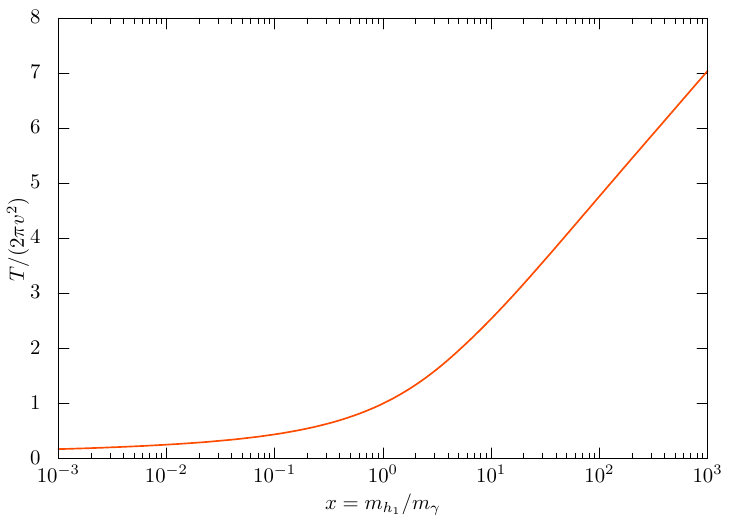}
\caption{String tension as 
a function of $x = m_{\fund_1}/m_\gamma$.}
\label{fig:string_tension}
\end{figure}

The string tension may be parameterized as
\begin{align}
    \mu = 2\pi \VH^2 f_T\qty(\frac{m_{\fund_1}}{m_\gamma})\ ,
\end{align}
where $m_{\fund_1}$ and $m_\gamma$ are the doublet Higgs mass 
and the massive U(1) gauge boson mass, respectively:
\begin{align}
    m_{\fund_1} = 2\sqrt{\Hlambda} \VH
    \ , \quad m_\gamma =\frac{1}{\sqrt{2}} g \VH\ .
\end{align}
\Cref{fig:string_tension} shows $f_T(x)$ calculated numerically.
It can be well approximated by
\begin{equation}
    f_T(x)=\frac{
    \begin{aligned}
        \Bigl[0.989951 &-0.266062\ln x + 0.0229062(\ln x)^2\\
        &+ x\qty( -0.0351507 -  0.0374918\ln x+0.0410306(\ln x)^2)\Bigr]
    \end{aligned}
    }{
    \begin{aligned}
        \Bigl[1 &-0.728749\ln x +0.195677(\ln x)^2\\
        &-0.0229062(\ln x)^3 + x(-0.0449826 + 0.0410306\ln x) \Bigr]
    \end{aligned}
    } \ .
\end{equation}
Note that the tension approaches
\begin{equation}
 \mu \to 
 2\pi v^2 \ln \frac{m_{\fund_1}}{m_\gamma} \ ,    
\end{equation}
for  $m_{\fund_1}/m_\gamma \gg 1$ and 
\begin{equation}
 \mu \to 
 2\pi v^2 
 \frac{1}{\ln [\modifiedat{JHEP2}{m_{\gamma}/m_{\fund_1}}]}
\end{equation}
for $m_{\fund_1}/m_\gamma \ll 1$ (see Ref.\,\cite{Yung:1999du}).

\subsection{Monopoles Connected by Cosmic String}
In the full theory, the gauge symmetry is completely broken:
\begin{align}
    \SU(2) \stackrel{V,\,v}{\longrightarrow}
    \mbox{nothing}\ .
\end{align}
Since $\pi_2(\SU(2))=\pi_1(\SU(2))=0$, no topological defects are allowed.

As we discussed above, however,
we expect monopoles as well as 
cosmic strings at each step of the hierarchical symmetry breaking.
In this subsection, we discuss how monopoles once appeared at the first symmetry breaking disappear at the next step.

To see the fate of a monopole, it is useful
to adopt the gauge consisting of  
two slightly overlapping charts covering the northern and southern hemispheres,
\begin{align}
\label{eq:NSchart}
    U_{N} &= 
    \left\{(r, \zenith,\azimuth)|0 \le \zenith \le {\pi}/{2} + \varepsilon
    ,\, r>R\right\}\ ,\\
    U_{S} &= 
    \left\{(r, \zenith,\azimuth)|{\pi}/{2}-\varepsilon \le \zenith \le \pi
    ,\, r>R\right\}\ .
\end{align}
Here, $\zenith$ is the 
zenith angle, $\varepsilon$ is a small positive number, and $R$ is 
some length scale satisfying 
$R\gg m_W^{-1}$.
Then, the monopole configuration \eqref{eq:heg} transforms as
\begin{align}
    \phi^a \sugen{a} &\to  \phi{}_{N,S}^a \sugen{a} = g_{N,S} \phi^a \sugen{a} g_{N,S}^\ct\ , \\
A^a_{i}\sugen{a} &\to  A_{N,S\,i}^a \sugen{a} = g_{N,S} A_{i}^a\sugen{a} g_{N,S}^\ct +
\frac{i}{g} g_{N,S} \partial_ig_{N,S}^\ct\ ,
\label{eq:localDP}
\end{align}
with
\begin{align}
\label{eq:combedgauge}
    &g_N =\left(
    \begin{array}{cc}
      c_{\zenith/2}  &  e^{-i\azimuth}s_{\zenith/2} \\
      - e^{i\azimuth}s_{\zenith/2}& c_{\zenith/2}
    \end{array}
    \right)\ ,
    ~~~~~~~ g_S =\left(
    \begin{array}{cc}
       e^{i\azimuth}c_{\zenith/2}  & s_{\zenith/2} \\
      -s_{\zenith/2}& e^{-i\azimuth}c_{\zenith/2}
    \end{array}
    \right)
\end{align}
in each chart.
We call this the combed gauge.

In the combed gauge, the asymptotic behavior of the monopole at $r \gg m_W^{-1}$ is given by%
\footnote{Here, we denote the gauge potential as a one-form.}
\begin{align}
\label{eq:phiN}
    \phi{}_N^a &\to \VG \delta^{a3}\ ,\qquad
    A^{a}_{N} \to \delta^{a3} (\cos\zenith -1) \dd{\azimuth}
\end{align}
in the $U_N$ chart and
\begin{align}
\label{eq:phiS}
    \phi{}^a_S  &\to \VG \delta^{a3}\ ,\qquad
    A^{a}_{S}\to \delta^{a3} 
    (\cos\zenith+1)  \dd{\azimuth}
\end{align}
in the $U_S$ chart. 
The other components
$A^{a}_{N,S}$ ($a=1,2$) vanish asymptotically for $r \gg m_W^{-1}$.

\begin{figure}[t]
\begin{center}
\begin{minipage}{.3\linewidth}
\includegraphics[width=.8\linewidth]{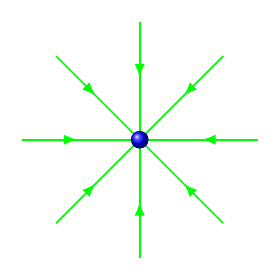}
\end{minipage}
 \begin{minipage}{.4\linewidth}
\hspace{2cm}
\includegraphics[width=.55\linewidth]{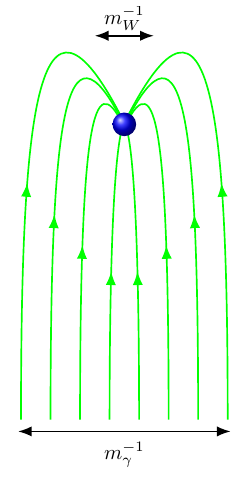}
 \end{minipage}
 \begin{minipage}{.6\linewidth}
\includegraphics[width=
\linewidth]{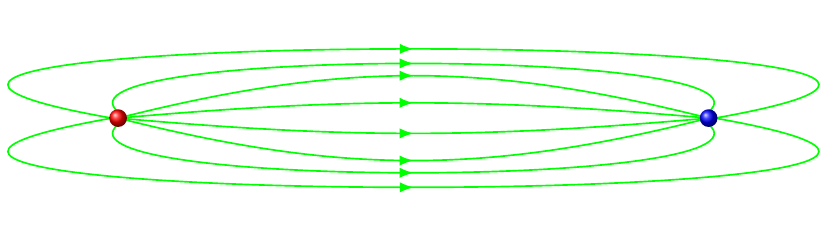}
 \end{minipage}
\end{center}
\caption{Schematic pictures of an isolated monopole, a monopole trapped inside a cosmic string and a monopole--antimonopole pair connected by cosmic string.}
\label{fig:trap}
\end{figure}

In the combed gauge,
the VEV of the adjoint scalar takes the same form as the vacuum in Eq.\,\eqref{eq:VEV1}, and hence, the U(1) gauge potential corresponds to $A_\mu^3$ far from the monopole. 
Due to the monopole, however,
$A^3_{N,S}$ in each chart are
connected at around the equator $\zenith\sim \pi/2$ by the non-trivial gauge transition function 
\begin{align}
    \label{eq:transition}
    t_{NS} = e^{2i\azimuth}\ ,
\end{align}
with which
\begin{align}
    A^{3}_{S}= A^{3}_{N}+ 2 \dd{\azimuth}\ .
\end{align}

Now we discuss the 
vacuum structure of $h_1$ around the monopole.
First, let us suppose that $h_1$ takes a constant expectation value $\VH$
in the northern hemisphere 
in the region far from the monopole, $r \gg m_W^{-1}$.
Then the U(1) magnetic flux
of $A^3$ is expelled from the northern hemisphere due to the Meissner effect, and hence,
the gauge potential in the northern hemisphere is trivial:
\begin{equation}
    A^{3}_{N} = 0
\end{equation}
for $r\gg m_W^{-1}$.
In the southern hemisphere 
below the overlap,
on the other hand, the doublet scalar and the gauge potential take the form\,\footnote{Note that the U$(1)$ charge of $h_1$ is $1/2$, which results in 
the cosmic string with the winding number $-1$ in the south\modifiedat{JHEP2}{ern} hemisphere.
Incidentally, \modifiedat{JHEP2}{if U(1) is broken by another SU(2) triplet instead of a doublet},
the same gauge transition function $t_{NS}$ of the 
monopole configuration
results in the bead configuration~\cite{Hindmarsh:1985xc,Everett:1986eh,Kibble:2015twa} (see also Refs.~\cite{Hiramatsu:2021kvu,Chitose:2023bnd}).
}
\begin{align}
\label{eq:phiS2}
    {h_1}_S &= e^{i\azimuth} h_1{}_N \ , \\
     A_{S}^3 &= \modifiedat{JHEP2}{2}\dd \azimuth \ ,
\end{align}
for $r\gg m_\gamma^{-1}$ due to the non-trivial transition function \eqref{eq:transition}.

As a result,
the trivial configuration of 
$h_1$
in the northern hemisphere
gives rise to a cosmic string 
configuration of $h_1$ with the winding number $n=-1$
in the southern hemisphere.
Importantly, the magnetic charge of the monopole coincides with the magnetic flux going through the cosmic string (see Eqs.\,\eqref{eq: monopole flux} and \eqref{eq: string flux}).
Therefore, at least one cosmic string attaches to a magnetic monopole produced at the first phase transition.
The other end of the string attaches to an
antimonopole 
(see Fig.~\ref{fig:trap}). The monopole and antimonopole connected by the string eventually annihilate and disappear from the Universe.

\section{String Breaking}
\label{sec:Ansatz}
In this section, we discuss
the breaking process of cosmic strings in our model.
They are spontaneously cut via a monopole-antimonopole creation inside, which is a tunneling process.
In the following, we compare two kinds of estimates of the bounce action:
the conventional one that neglect the soliton sizes
and the one based on the Ans\"atze proposed in
Ref.\,\cite{Shifman:2002yi}
which takes them into account.

\subsection{Breaking of Infinitely Thin String}
\label{sec: Thin String}
\begin{figure}
    \centering
\includegraphics[width=0.5\textwidth]{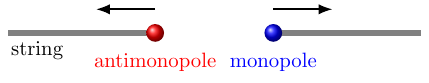}
    \caption{Cosmic string breaking through monopole-antimonopole pair production.\label{fig:pairproduction}}
\end{figure}

Here, we briefly review the conventional estimate of the string breaking rate in the infinitely thin cosmic string limit~\cite{Preskill:1992ck}, which we call the Preskill-Vilenkin approximation.
As discussed in the previous section, a cosmic string can end with a monopole in our setup.
This means that a cosmic string can be cut by a monopole-antimonopole pair production 
as a tunneling process (see \cref{fig:pairproduction}).

\begin{figure}
    \centering
    \includegraphics[width=0.45\textwidth]{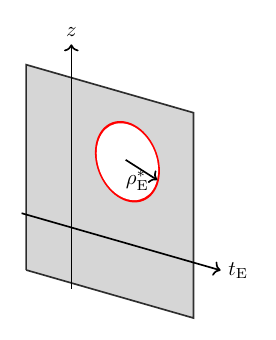}
    \caption{Bubble on the Euclidean string worldsheet. The circular monopole worldline is drawn in red.\label{fig:hole}}
\end{figure}

To calculate the tunneling factor, we use the Euclidean path integral method 
($t = -i\tE$). 
In the infinitely thin string limit, the string breaking process may be regarded as a vacuum decay in 1+1 dimensions.
That is, the string corresponds to the false vacuum and the absence of string corresponds to the true vacuum.
The monopole plays the role of the domain wall separating the two vacua (see \cref{fig:hole}).

In the Minkowski space, 
the cosmic string is invariant with respect to Lorentz boosts along the string, on which we place the $z(=x^3)$ axis.
Lorentz boosts in the $(t,z)$ plane correspond to rotations in the Euclidean $(\tE,z)$ plane.%
\footnote{The magnetic field along the cosmic string corresponds to $F^{\mathrm{U(1)}12}\neq 0$, which is invariant under either Lorentz boosts 
in the $(t,z)$ plane
or SO(2) rotations in the $(\tE,z)$ plane.}
We assume that the bounce solution preserves this symmetry, and hence, 
the domain wall separating the two vacua (i.e. monopole worldline) is a circle on 
the $(\tE,z)$ plane.

The bounce action of the bubble is given by,
\begin{align}
    S_B&=m_M\int_{\text{worldline}} \dd{x}- \mu\int_{\text{hole}}\dd[2]{S} \\
    &=2 \pi \brad m_M - \pi \brad^2 \mu\ ,
\end{align}
where $\mu$ is the string tension, $m_M$ is the monopole mass and $\brad$ is the radius of the monopole worldline.

Maximizing this with respect to $\brad$, we obtain
\begin{align}
    \brad&=\frac{m_M}{\mu}\ , \\
    S_B^{(\mathrm{thin})}&=\frac{\pi m_M^2}{\mu}=:\pi\kappa\ .\label{eq:Preskill}
\end{align}
The string breaking rate per unit length is given by
\begin{equation}
    \Gamma \simeq \frac{\mu}{2\pi} e^{-S_B^{(\mathrm{thin})}}\ .
\end{equation}
For the the prefactor, 
see e.g. Ref.~\cite{Kiselev:1975eq}.
For reference, the recent PTA
data are compatible with
the GWs from metastable cosmic strings with $\sqrt{\kappa} \sim 8$.

\subsection{Primitive Ansatz of Unwinding Process}
\begin{figure}
    \centering    \includegraphics[width=0.45\textwidth]{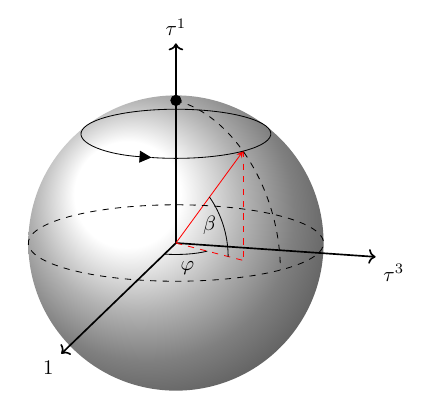}
    \caption{Unwinding of $U$ in the subspace of $\SU(2)$ without $\pauli{2}$ component.
    The matrix $U$ winds around the sphere at $\unw=0$, i.e., $U=e^{-i\varphi \pauli{3}}$, while $U=i\pauli{1}=\const$ at $\unw=\pi/2$.
    }
    \label{fig:unwindingU}
\end{figure}
Following Ref.~\cite{Shifman:2002yi},
we parameterize the unwinding of a cosmic string with $\unw$
by introducing two Ans\"atze.%
\footnote{The unwinding parameter $\unw$ in this paper 
corresponds to $\zenith$ in Ref.\,\cite{Shifman:2002yi}. 
In this paper, $\zenith$ is reserved for the zenith angle.}
The first one 
called the ``primitive'' Ansatz in Ref.~\cite{Shifman:2002yi} is 
as follows:%
\footnote{The other Ansatz which is called ``improved" Ansatz is discussed in \cref{sec:improved}.}
\begin{align}\label{eq:primitive}
    \fund(x)&=U\pmqty{\higgsprofile_\unw(\rho) \\ 0}\ ,\\
    A_\rho(x)&=0\ ,\\
    A_\az(x)&=iU(\partial_\az U^\ct) (1-f_\unw(\rho))\ ,\\
    \phi(x)&=VU\frac{\pauli{3}}{2}U^\ct + \Delta\phi\ , 
    \label{eq:phi}
\end{align}
where
\begin{equation}
    U=e^{-i\azimuth\pauli{3}}\cos\unw + i\pauli{1}\sin\unw
\end{equation}
and
\begin{equation}
    \Delta\phi=\Phi_\unw(\rho)\bqty{\frac{\pauli{1}}{2}\sin\az-\frac{\pauli{2}}{2}\cos\az}\ .
\end{equation}
Here, we adopted the cylindrical coordinates $(t,z,\rho,\az)$ with the $z$ axis along the string.
The boundary conditions for the profile functions are
\begin{gather}
    \higgsprofile_\unw(\rho=0)=0\ ,\qquad f_\unw(\rho=0)=1\ , \label{eq:originBC}\\
    \higgsprofile_\unw(\rho\to\infty)=v\  , \qquad f_\unw(\rho\to\infty)=0\ , \label{eq:infinitBC}
\end{gather}
and
\begin{equation} \label{eq:Phiboundary}
    \Phi_\unw (\rho=0)=V\sin 2\unw\ ,\ \qquad \Phi_\unw(\rho\to\infty)=0\ .
\end{equation}
The former of \cref{eq:Phiboundary} ensures $\phi$ is well-defined at $\rho=0$.
For the static configuration, we take $A_t(x)=A_z(x)=0$. We will come back to this point later.

\begin{figure}

 	\subcaptionbox{Configuration of $\fund_1$\label{fig:phase1}}
	{\includegraphics[ width=0.49\textwidth]{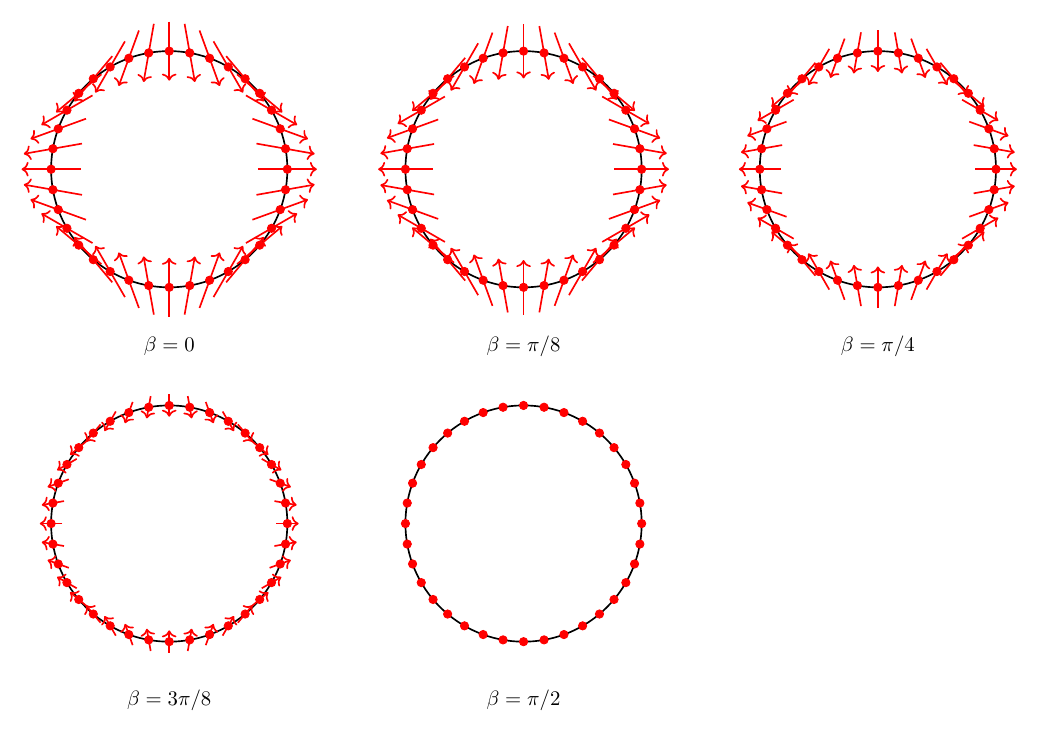}}
\hspace{0.1\columnwidth}
 	\subcaptionbox{Configuration of $\phi^a$ \label{fig:phase2}}
	{\includegraphics[ width=0.3\textwidth]{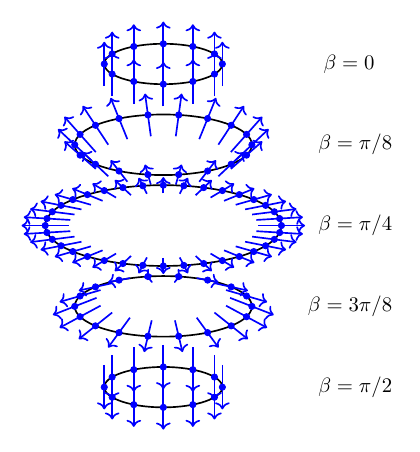}}
    \caption{\modifiedat{JHEP2}{(a): The winding component 
    $\fund_1(x)\propto e^{-i\varphi}$
    projected on to the $(x_1,x_2)$ plane, which is perpendicular to the string. 
    The real and imaginary axis of $h_1$ are projected to the $x_1$ and $x_2$ axis, respectively.
    The string is placed at the center of the circle.
    The direction of the winding component rotates clockwise 
    as we circle around the cosmic string.
    The winding component vanishes at $\unw = \pi/2$.
    (b): The directions of $\vec{\phi}(x)$ displayed as spatial vectors.
    The $\pauli{i}$ component is identified with the $i$'th spatial component.
    Here, we dropped $\Delta \phi$, which is subdominant at large $\rho$.
    When stacked vertically, they form the hedgehog structure.
    }
}
\end{figure}

The angle $\unw$ parameterizes
the winding of $h_1$
through the winding matrix $U\in \SU(2)$.
\Cref{fig:unwindingU} shows $U$ in the $S^2$ subspace of SU(2) without the $\pauli{2}$ component.
At $\unw = 0$, $U= e^{-i\varphi \pauli{3}}$, and hence,
it rotates in the $\1_2$--$\pauli{3}$ plane as $\az$ goes from 0 to $2\pi$.
The configuration at $\unw = 0$ corresponds to 
the cosmic string with the winding number $-1$.
Intermediate $\unw$
describes a configuration with a smaller winding 
component
(see \cref{fig:phase1}).
At $\unw = \pi/2$, $U=i\pauli{1}=\const$
and hence, $\fund_1$ no longer winds.
In this way, the unwinding parameter $\unw$ interpolates
the cosmic string $\unw = 0$
and the ``vacuum'' $\unw = \pi/2$.%
\footnote{Technically, $\higgsprofile_\unw$ cannot reach the vacuum even for $\unw=\pi/2$, as $\higgsprofile_\unw(0)=0$ is imposed. Thus, the string tension is nonzero for all $\unw$.}

Also, \cref{fig:phase2}
shows the 
directions of $\phi^a$ 
on the ($x_1,x_2$) plane.
We omitted
$\Delta \phi$ which is subdominant at large $\rho$.%
\footnote{In the figure, 
we show 
$e^{i(\pi/2) \pauli{3}/2}\phi e^{-i(\pi/2) \pauli{3}/2}$ instead of 
$\phi$
so that the hedgehog structure is apparent.
}
The figure shows that $\phi^a$ flips 
as $\unw$ changes from
$0$ to $\pi/2$.
If one stacks \cref{fig:phase2} vertically, the arrows form the hedgehog shape, indicating the formation of a monopole.

For later use, we also display the primitive Ansatz in the singular gauge:\,\footnote{In the singular gauge, $\varphi$ dependence persists even for $\rho \to 0$.}
\begin{align}\label{eq:primitiveSingular}
    \fund(x)&=\pmqty{\higgsprofile_\unw(\rho) \\ 0}\ ,\\
    A_\rho(x)&=0\ , \\
    A_\az(x)&=-i(\partial_\az U^\ct) U f_\unw(\rho)\ ,\\ \label{eq:primitiveSingularAaz}
    \phi(x)&=V\frac{\pauli{3}}{2} + U^\ct (\Delta\phi) U\ .
\end{align}
Substituting the Ansatz into the Lagrangian \eqref{eq:lagrangian} yields the string tension for a given $\unw$,
\begin{multline} \label{eq:primitiveT}
    T(\unw)=2\pi \int_0^\infty \rho\dd{\rho}\Biggl\{\frac{2}{g^2}\frac{(\delrho f_\unw)^2}{\rho^2}\cos^2\unw +(\delrho \higgsprofile_\unw)^2+\frac{f_\unw^2}{\rho^2}\higgsprofile_\unw^2 \cos^2\unw\\
    +\frac{1}{2}(\delrho \Phi_\unw)^{2}+\frac{1}{2\rho^2}\bqty{\Phi_\unw(\cos 2\unw-2f_\unw \cos^2\unw)+Vf_\unw \sin 2\unw}^2\\
    +\Hlambda(\higgsprofile_\unw^2-v^2)+\Vlambda\bqty{\Phi_\unw^2-2V\Phi_\unw \sin 2\unw}^2+\frac{\gamma}{4}\Phi_\unw^2 \higgsprofile_\unw^2 \Biggr\}\ .
\end{multline}
For a given $\unw$, the profile functions $f_\unw(\rho)$, $\higgsprofile_\unw(\rho)$ and $\Phi_\unw(\rho)$ are obtained by minimizing $T(\unw)$ numerically.

\begin{figure}
    \centering
   \includegraphics{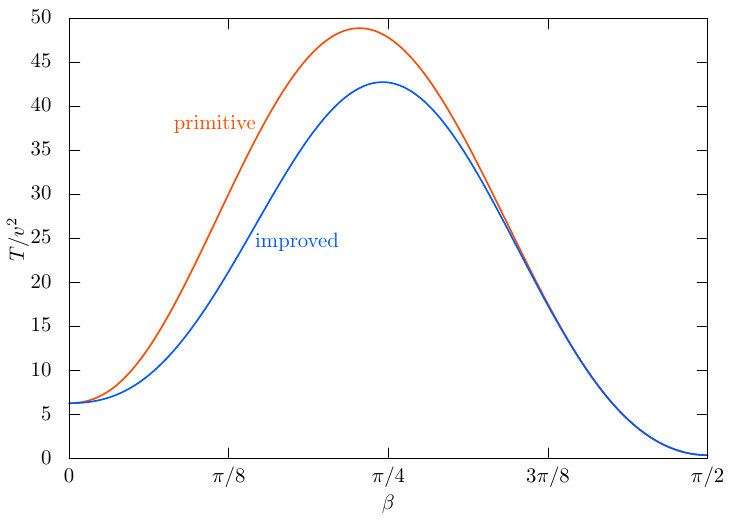}
      \caption{The string tension for a sample parameter set, $g=1$,
     $m_{\fund_1} = m_\gamma$, $m_W = \adjmass = m_{\fund_2} =5m_\gamma$.
     We show the string tension
     for the primitive (red)
     and improved (blue)
     Ans\"atze.
    The true vacuum $\unw=\pi/2$ has lower tension than the ANO string $\unw=0$.}
    \label{fig:betatension}
\end{figure}

\Cref{fig:betatension} shows $T(\unw)$ for a sample parameter set,
$g=1$, $m_{\fund_1} = m_\gamma$, $m_W = \adjmass = m_{\fund_2} = 5m_\gamma$.
The red line shows the string tension for the primitive Ansatz and the blue line shows that for the improved Ansatz introduced below.
The ``true'' vacuum ($\unw=\pi/2$) has lower tension than the ANO string ($\unw=0$), as it should.

Incidentally, we have checked 
that $\unw = 0$
is indeed a local minimum of the tension $T(\unw)$
for $m_W/m_\gamma \gtrsim 0.8$ 
for $m_W=\adjmass = m_{\fund_2}$ and $m_{\fund_1}=m_\gamma$.
Armed with this observation,
we assume that 
the string configuration is 
classically stable even for $m_W/m_\gamma = O(1)$,
while the classical stability is 
obvious for $m_W/m_\gamma \gg 1$.%
\footnote{To show the classical stability of the string configuration, we need to show that no tachyonic fluctuation exists around the string configuration with $\unw = 0$, which will be studied in a future work.}

\subsection{Effective Description}

To describe the string breaking, 
we promote the unwinding parameter 
$\unw$ to a collective coordinate 
for the unstable mode
on the string worldsheet 
and construct an effective 
1+1 dimensional theory of $\unw(t,z)$.%
\footnote{This formalism usually assumes $\unw$ to depend adiabatically on the worldsheet coordinates. 
In the string breaking process, 
the worldsheet coordinate dependence 
of $\unw$ is no more adiabatic.
Nevertheless, we use the effective $\unw(t,z)$ theory to find the 
path which connects the string configuration 
and the trivial vacuum.
The bounce action estimated in this way
should give an upper limit 
on the bounce action of the string breaking.
}
In this effective theory, string breaking 
is described as a tunneling process of $\unw$
through the potential barrier, i.e.,  $T(\unw)$.

Once we introduce $(t, z)$ dependence of $\unw$, the field strength
$F_{nj}$ ($n=t,z$) become singular 
at $\rho \to 0$ for $A_{t,z} = 0$. 
To avoid this singularity, 
we need to introduce an additional 
profile function $a_\unw(\rho)$ with which,\,\footnote{We could have defined $A_\mu=iU\partial_\mu{U^{-1}}(1-f_\unw)$ for all components and let the derivative act also on $\unw(t,z)$. 
Nevertheless, 
the choice in \cref{eq:An}
results in a smaller kinetic term, and hence, bounce action.}
\begin{equation}
\label{eq:An}
    A_n(x)=-2(\partial_n \unw(t,z))\pqty{\frac{\pauli{1}}{2}\cos\az-\frac{\pauli{2}}{2}\sin\az}a_{\unw}(\rho)\ , \quad (n=t,z)\ ,
\end{equation}
with $a_\unw(0)=1$.
Here, we again take the singular gauge.
The finiteness of the action also requires $a_\unw(\infty)=0$.
Note also that $A_n$ in \cref{eq:An} do not affect the estimate of $T(\unw)$ discussed in the previous section.

\subsection{Bounce Action}
\begin{figure}
    \centering
       \includegraphics[]{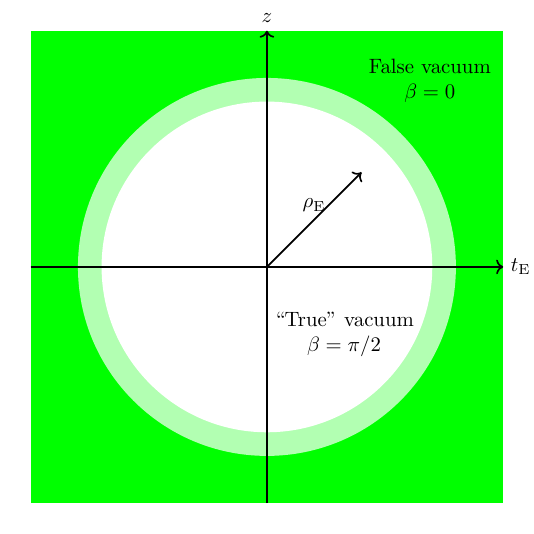}
    \caption{The bubble configuration on the Euclidean string worldsheet. The SO(2) symmetry allows the bounce solution to be parameterized by a single coordinate $\rhoE$.}
    \label{fig:bubble}
\end{figure}

\begin{figure}
    \centering
    \includegraphics{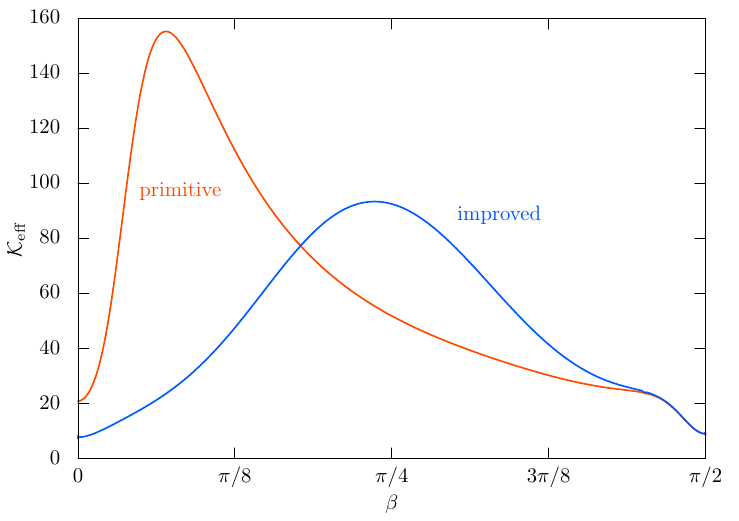}
    \caption{The coefficient $\Keff(\unw)$ of the kinetic term for the same parameter set as \cref{fig:betatension}.
    }
    \label{fig:betaKeff}
\end{figure}
Once we have constructed an effective two dimensional theory of $\unw(t,z)$ on the string world sheet, the string breaking process 
can be regarded as a bubble formation of 
the ``true'' vacuum with $\unw \simeq \pi/2$.
To calculate the bounce action of the bubble formation, we again consider the Euclidean action as in \cref{sec: 
Thin String}.
The Euclidean action takes the form
\begin{equation} \label{eq:Ha}
    \Seucl=\int \dd{t}_\mathrm{E}\dd{z} \dd{\rho} \bqty{\frac{1}{2}(\partial_n \unw(\tE,z))^2 \hamiltonian_\unw(a_\unw(\rho),\delrho a_\unw(\rho)) + (\text{$a$ independent})}\ ,
\end{equation}
where $n=(\tE,z)$.
The azimuthal angle $\az$ of 
the cylindrical coordinate 
around the cosmic string has already been integrated out and the Jacobian is absorbed to the integrand.
The prefactor of the kinetic term of $\pqty{\partial_n \unw}^2$, $\mathcal{H}_\unw$, is given 
in \cref{eq:Hprimitive}.

Note that \cref{eq:Ha} allows us to determine $a_\unw(\rho)$ beforehand to find the bounce solution of $\unw(\tE,z)$.
Since the bubble must minimize the action except for the direction of the bubble size, $a_\unw$ is determined to minimize $\int \dd{\rho} \hamiltonian_\unw(a_\unw(\rho),\delrho a_\unw(\rho))$ for each fixed $\unw$.

As discussed in \cref{sec: Thin String}, the string configuration is invariant under the Lorentz boost in $(t,z)$ plane in
the Minkowski space, we assume that the bubble configuration is SO(2) invariant 
in $(\tE,z)$ plane.
The bubble configuration for the tunneling
starts from $\unw = 0$ and reaches to $\unw \simeq \pi/2$ (see \cref{fig:bubble}).
Setting the origin of $(\tE,z)$ plane to be the bubble center and defining $\rhoE\defby \sqrt{\tE^2+z^2}$, we have
\begin{equation}
    \Seucl=2\pi \int \rhoE\dd{\rhoE} \int \dd{\rho}\bqty{\frac{1}{2}K(\unw(\rhoE),\rho)(\delrhoE \unw(\rhoE))^2+V(\unw(\rhoE),\rho)}\ .
\end{equation}
The explicit forms of $K$ and $V$ are given in \cref{eq:primitiveK} and \cref{eq:primitiveV}, respectively.
Carrying out the integration by $\rho$, we obtain
\begin{equation}\label{eq:euclideanAction}
    \Seucl=2\pi \int \rhoE \dd{\rhoE}\bqty{\frac{1}{2}\Keff(\unw(\rhoE))
    (\delrhoE \unw(\rhoE))^2+T(\unw(\rhoE))}\ ,
\end{equation}
where $\Keff(\unw)\defby \int \dd{\rho}K(\unw,\rho)$ and $T(\unw)= \int \dd{\rho}V(\unw,\rho)$ is the tension in \cref{eq:primitiveT}.
\Cref{fig:betaKeff} shows $\Keff(\unw)$ for the same parameter set as \cref{fig:betatension}.

The Euclidean equation of motion (EOM) is
\begin{equation}\label{eq:euclideanEoM}
    \Keff(\unw(\rhoE))\unw''(\rhoE)=\Veff'(\unw(\rhoE))-\frac{1}{2}\Keff'(\unw(\rhoE))\unw'(\rhoE)^2-\frac{1}{\rhoE}\Keff(\unw(\rhoE))\unw'(\rhoE)\ 
    ,
\end{equation}
where a prime denotes the derivative with respect to the argument.
By bounce solution which satisfies
\begin{align}
\label{eq:bounceBC}
    \unw(0) \simeq \frac{\pi}{2}\ ,\ \quad \unw'(0)=0 \ , \quad \unw(\rho_E\to \infty) =0\ ,
\end{align}
we obtain the bounce action 
of the unwinding process,
\begin{align}
   S_{B}\defby \eval{\Seucl}_\text{bounce}-\eval{\Seucl}_{\unw=0}\ .
\end{align}

\subsubsection*{Thin-wall approximation}
In our analysis, we solve the bounce equation numerically, where the effective 
$\mathcal{K}_\mathrm{eff}(\unw)$ and $T(\unw)$ are also estimated numerically.
In the limit of $\VG\gg \VH$, however,
the $T(0)-T(\pi/2)$ is much smaller than
the typical height of $T(\unw)$.
In such a case, the so-called thin-wall approximation provides a good estimate 
of the action.
For later purpose, we derive the bounce 
action in this approximation.

The EOM~\eqref{eq:euclideanEoM} can be rewritten as
\begin{equation}\label{eq:energyConservation}
    \dv{\rhoE} \bqty{\frac{1}{2}\Keff \unw'^2 - \Veff}=\frac{1}{\rhoE}\Keff \unw'\ .
\end{equation}
For $\VG\gg \VH$, the critical bubble radius $\brad$ should be large compared to the bubble thickness.
Thus, around the bubble wall, the right hand side of \eqref{eq:energyConservation} effectively vanishes and the quantity in the brackets is conserved.
Since the boundary condition at $\unw=0$ is $\unw'(\rhoE)=0$,
\begin{equation}
    \frac{1}{2}\Keff(\unw)\unw'(\rhoE)^2-\Veff(\unw)=-\Veff(0)\ .
\end{equation}

We evaluate the bounce action by separating it to three parts.
The region outside the bubble makes no contribution as $\unw'\approx 0$ and $\Veff(\unw)-\Veff(0)=0$.
On the bubble wall,
\begin{align}
2\pi\int_{\text{wall}} \rhoE\dd{\rhoE} \bqty{\frac{1}{2}\Keff(\unw)\unw'^2+\Veff(\unw)-\Veff(0)} &= 2\pi \brad\int_{\text{wall}}\dd{\rhoE} \bqty{\Keff(\unw)\unw'^2} \\
&= 2\pi\brad \int_{\text{wall}}\dd{\unw} \sqrt{2\Keff(\unw)(\Veff(\unw)-\Veff(0))}\\
&\approx 2\pi\brad \int_{0}^{\pi/2}\dd{\unw} \sqrt{2\Keff(\unw)(\Veff(\unw)-\Veff(0))}\\
&=: 2\pi \brad \meff\ .
\end{align}
The contribution from the potential difference inside the wall is
\begin{equation}
    -\pi\brad^2\pqty{\Veff(0)-\Veff\pqty{\frac{\pi}{2}}}\ .
\end{equation}
In total,
\begin{align}
    S =  2\pi \brad \meff  -\pi\brad^2\pqty{\Veff(0)-\Veff\pqty{{\pi}/{2}}} \ .
\end{align}
Maximizing this with respect to $\brad$, we obtain
\begin{align}
\label{eq:thinWALL}
   \brad&=\frac{\meff}{\Veff(0)-\Veff(\pi/2)}\ ,\\    S_B &= \eval{\Seucl}_{\text{bounce}}-\eval{\Seucl}_{\unw=0}=\frac{\pi \meff^2}{\Veff(0)-\Veff(\pi/2)}\ .
\end{align}
Unlike ordinary vacuum decays, $S_B$ depends not only on the potential 
barrier 
$T(\unw)$ but also on the $\unw$-dependent mass $\Keff$.

By comparing with the bounce action in \cref{eq:Preskill}, 
the monopole mass $m_M$ is replaced by $\meff$.%
\footnote{Both $\Keff$ and $T$ depend on $\rhoE$ only through 
$\unw(\rhoE)$.}
As we will see later, they are found to be numerically close for 
$V \gg v$.
Thus, the primitive Ansatz describes 
the string breaking process via the (excited) monopole-antimonopole pair creation.

\subsection{Improved Ansatz}
\label{sec:improved}
The gauge field for the primitive Ansatz in the 
singular gauge \eqref{eq:primitiveSingularAaz} may be written more explicitly as
\begin{equation}
    A_\az = \bqty{2\cos^2\unw \frac{\pauli{3}}{2}-\sin 2\unw\pqty{\frac{\pauli{1}}{2}\sin\az+\frac{\pauli{2}}{2}\cos\az}}f_\unw(\rho)\ .
\end{equation}
Since it depends on $\rho$ only through $f_\unw$, the monopole and the string must share their radial variation size.
Authors of Ref.~\cite{Shifman:2002yi} argued that it leads to an overestimate of $T(\unw)$ and the bounce action, as the monopole, which has the natural size $\sim \VG^{-1}$, must spread over the width of the string $\sim v^{-1}$.

With this in mind, they introduced the improved Ansatz in the singular gauge:
\begin{equation}
    A_\az = 2\cos^2\unw \frac{\pauli{3}}{2}\fgamma-\sin 2\unw\pqty{\frac{\pauli{1}}{2}\sin\az+\frac{\pauli{2}}{2}\cos\az}\fW\ .
\end{equation}
We have two separate profiles for the $\pauli{3}$ component and for the others.
Since the former is responsible for the string formation, $\fgamma$ is expected to spread over $\rho\sim v^{-1}$ while $\fW$ is responsible for the monopole and should have the length scale $\sim \VG^{-1}$.

Requiring $A$ to be regular at $\rho=0$ in the regular gauge, we obtain the boundary conditions
\begin{alignat}{2}
    \fgamma(0)&=\fW(0)&&=1\ ,\\
    \fgamma(\infty)&=\fW(\infty)&&=0\ .
\end{alignat}

Everything else is the same as the primitive Ansatz.
Corresponding $\hamiltonian_\unw$, $K(\unw)$ and $V(\unw)$ are written down in \cref{sec:longeqs}.
In our numerical analysis, we compare the primitive and the improved Ans\"atze.

In Ref.~\cite{Shifman:2002yi},
the authors argued that 
the primitive Ansatz 
overestimates the string tension
by a factor 
of $\ln(\adjmass/m_\gamma)$
for $\VG \gg \VH$ compared with the improved Ansatz.
As can be seen in \cref{fig:betatension}, the modification indeed reduces the string tension.
However, even with large hierarchy, the improvement is not as drastic as was claimed.
Contrary to their expectations, the monopole does not seem to spread to the scale $\sim v^{-1}$ even in the primitive Ansatz; rather, the string shrinks to keep the monopole small.
Thus, the logarithmic enhancement is absent from both the Ans\"atze.

\section{Numerical Results}
\label{sec:Numerical}

\begin{figure}[tb]
\centering
\begin{subfigure}{0.50\linewidth}
    \centering
    \includegraphics[width=\linewidth]{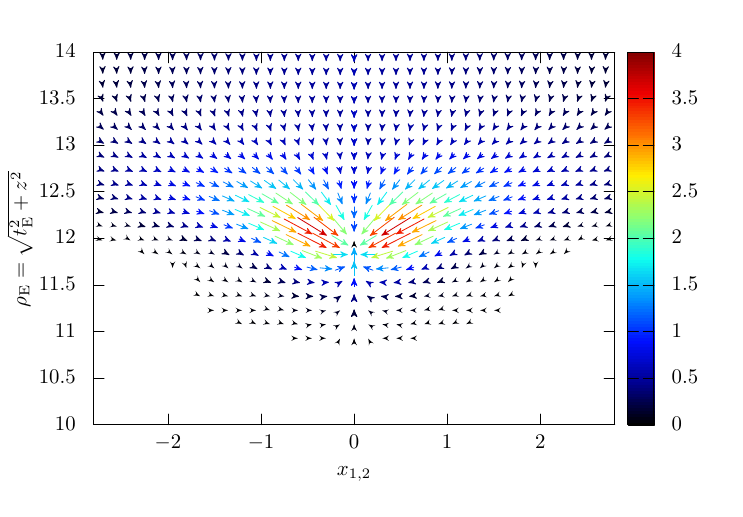}    
    \caption{Primitive Ansatz}
    \label{fig:monopole_config_primitive}
\end{subfigure}%
\hfill%
\begin{subfigure}{0.50\linewidth}
    \centering
    \includegraphics[width=\linewidth]{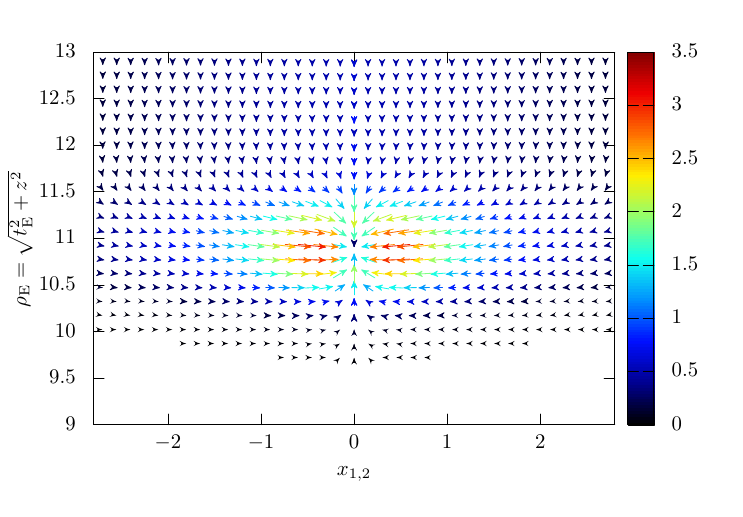}
    \caption{Improved Ansatz}
    \label{fig:monopole_config_improved}
\end{subfigure}
\caption{
The magnetic field around the 
(excited) monopole 
created at the string breaking 
for two Ans\"atze.
The size and color of the arrows represent the field strength.
Here, 
we take 
 $g=1$,
     $m_{\fund_1} = m_\gamma$, $m_W = \adjmass = m_{\fund_2} =5m_\gamma$ with $\VH=1$.
}
\label{fig:monopole_config}
\end{figure}

Our numerical analysis procedure is summarized below.
\begin{enumerate}
    \item Determine profile functions $\higgsprofile_\unw(\rho)$, $f^{(\gamma,W)}_\unw(\rho)$, and $\adjprofile_\unw(\rho)$ satisfying the boundary conditions in \cref{eq:originBC,eq:infinitBC,eq:Phiboundary} by
    minimizing $T(\unw)$ for each $\unw$.
    \item 
    Turn on the $(t,z)$ dependence of $\unw$.
    Substitute $\higgsprofile_\unw(\rho)$, $f^{(\gamma,W)}_\unw(\rho)$, and $\adjprofile_\unw(\rho)$  
    and their derivatives into $\mathcal{H}_\unw$ in \cref{eq:Ha}.
    Determine $a_\unw(\rho)$
    satisfying the boundary condition below \cref{eq:An} by minimizing $\int \dd{\rho} \mathcal{H}_\unw$.
    \item Substitute the Ansatz into the action and obtain the 2D effective theory on $(t,z)$ by integrating over $(\rho,\az)$.
    \item Obtain the bounce solution satisfying the boundary conditions \eqref{eq:bounceBC}
    by assuming the $\operatorname{SO}(2)$ symmetry in the $(\tE,z)$ plane.
\end{enumerate}

In the following, we 
rescale the scalar fields as 
$\hat{\phi}\defby g\phi$ and $\hat{\fund}\defby g\fund$ and 
parameterize the
dimensionless coupling
constants $\Vlambda$, ${\lambda}$ and $\gamma$
in terms of the ratios of the mass parameters:
\begin{align}
\frac{\lambda}
{g^2} =  \frac{m_{\fund_1}^2} {8m_\gamma^2} \ ,  \qquad
\frac{\Vlambda}
{g^2} =  \frac{\adjmass^2} {8m_W^2} \ , \qquad 
\frac{\gamma}{g^2} = \frac{m_{\fund_2}^2}{m_W^2}  \ .
\end{align}
With this parameterization, the bounce action 
is proportional to $g^{-2}$, which can be factored out.
Note that the same is true for the bounce action for the Preskill-Vilenkin approximation in \cref{eq:Preskill}.%
\footnote{Due to the rescaling of $\fund$ and $\phi$, 
the boundary conditions of the profile functions are changed to $\hat{\higgsprofile}_\unw(\rho\to \infty) = \sqrt{2}m_\gamma$ and 
$\hat{\adjprofile}_\unw(\rho= 0) = m_W \sin 2\unw$.}

\Cref{fig:monopole_config} shows the magnetic field around the (excited) monopole created at the 
string breaking for the two Ans\"atze.
The magnetic monopole is at around $x_{1,2} = 0$, while 
the cosmic string extends to the top. 
The magnetic flux of the cosmic string flows into the monopole.
The size of the monopole (the support of $\adjprofile_\unw$) is $\sim O(m_W^{-1})$ and much smaller than the string width $\sim O(m_\gamma^{-1})$.
Note that the magnetic flux disappears at $x_{1,2}\simeq 0$ 
since $\phi^a \to 0$ at $x_{1,2}\to 0$
for $\unw = \pi/4$.
In the improved Ansatz, the separation of $f^W$ and $f^\gamma$ allows the magnetic field
along the cosmic string to be frozen until just above the monopole (\cref{fig:monopole_config_improved}).

\begin{figure}[tb]
\centering
\includegraphics[width=0.48\linewidth]{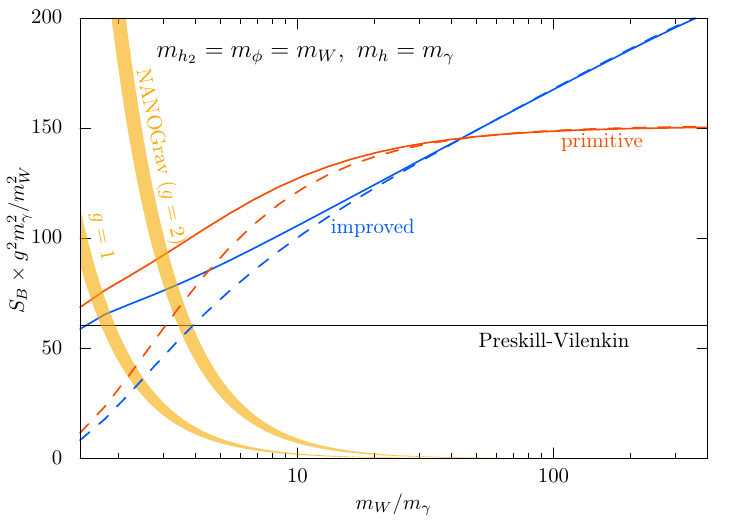}    \includegraphics[width=0.48\linewidth]{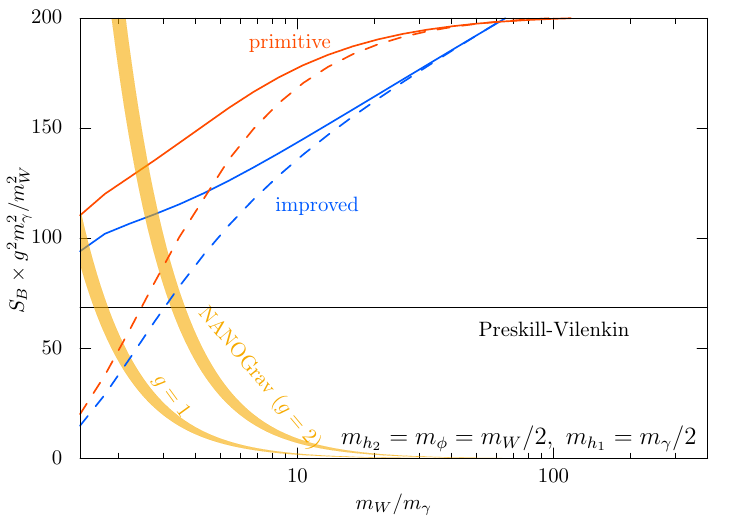}    \includegraphics[width=0.48\linewidth]{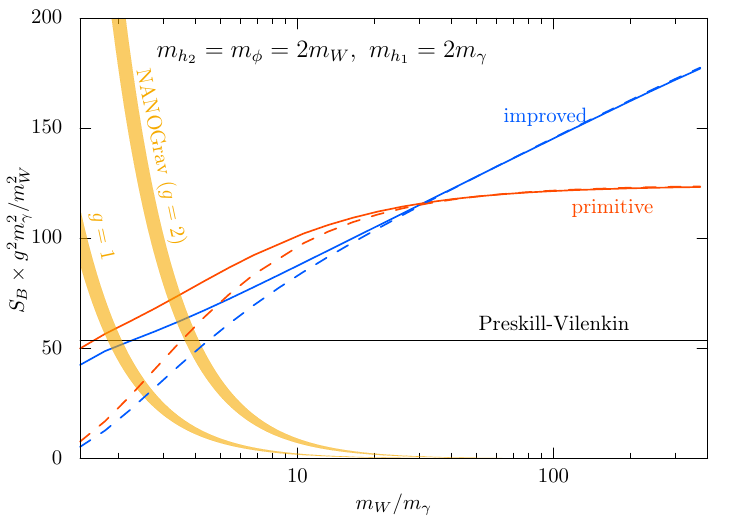}
 \includegraphics[width=0.48\linewidth]{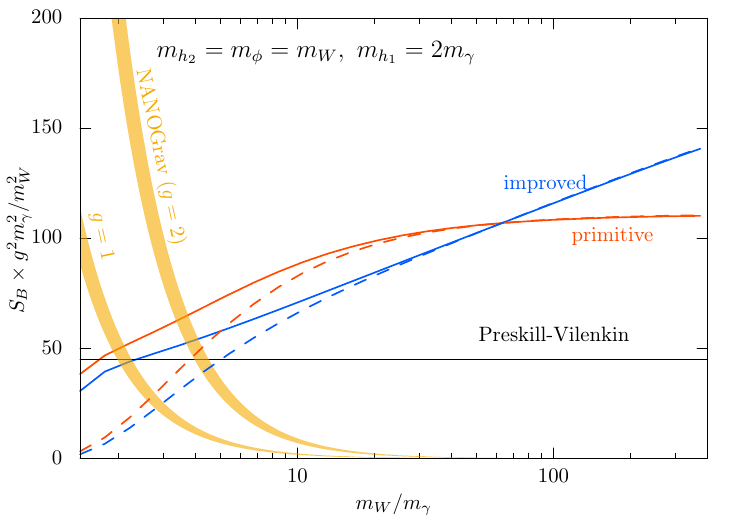}
\caption{The bounce action 
normalized by $g^2 m_\gamma^2/m_W^2$ 
for the primitive (red)
and improved (blue)
Ans\"atze.
The dashed-lines show the thin-wall approximation of 
each Ansatz in \eqref{eq:thinWALL}. 
We also show the bounce action in the infinitely thin string limit $S_B^{\mathrm{(thin)}}$ (black).
The yellow bands show 
the bounce action compatible with 
the PTA data, $S_B = \pi \kappa$,
with $\sqrt{\kappa}=7.5$--$8.5$ 
for $g=1$ and $g=2$.
}
\label{fig:bounce-susy}
\end{figure}

\begin{figure}
    \centering
    \begin{subfigure}[t]{0.48\linewidth}
        \centering
        \includegraphics[width=\linewidth]{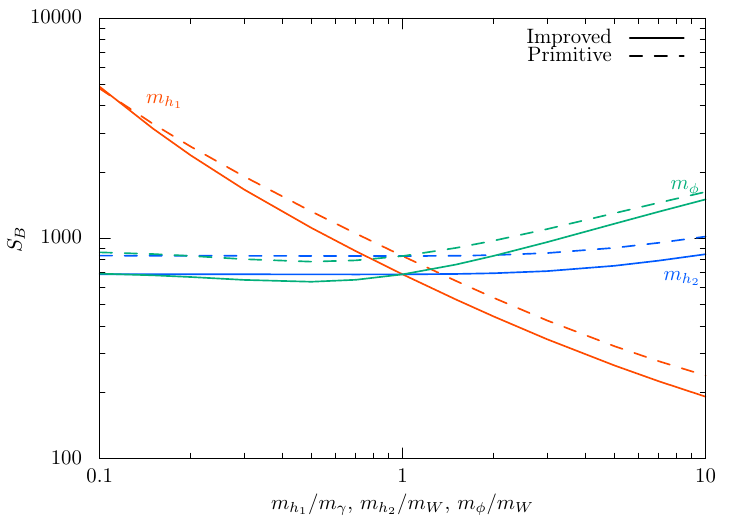}
        \caption{}
        \label{fig:mass_dependence}
    \end{subfigure}%
    \hfill%
    \begin{subfigure}[t]{0.48\linewidth}
        \centering
        \includegraphics[width=\linewidth]{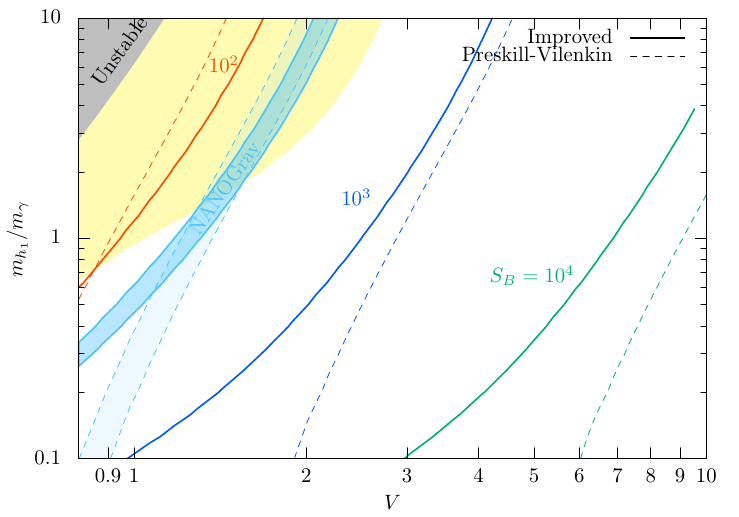}
        \caption{}
        \label{fig:ratio_v_contour}
    \end{subfigure}
    \caption{(a) The mass ratio dependence of $S_B$ with $g=1$ and $m_W/m_\gamma=5$. The solid (dashed) lines correspond to the improved (primitive) Ansatz. The bounce action is shown as a function of one of $m_{\fund_1}/m_\gamma$, $m_{\fund_2}/m_W$, $\adjmass/m_W$, with the other two fixed to unity.
    (b) The bounce action as a function of $\VG$ (with $\VH=1$) and $m_{\fund_1}/m_\gamma$, with $g=1$ and $\adjmass=m_{\fund_2}=m_W$, calculated by the improved Ansatz.
        The dashed lines show the bounce action by the Preskill-Vilenkin approximation.
          In the yellow shaded region, 
        the bounce action in the improved Ansatz is smaller than that in the Preskill-Vilenkin approximation.
        The cyan band corresponds to the PTA data, $\sqrt{\kappa}\simeq\numrange{7.5}{8.5}$.
        The lighter cyan band shows the same in the Preskill-Vilenkin approximation.
      }
    \label{fig:parameter-dependence}
\end{figure}

\Cref{fig:bounce-susy} shows the bounce actions
normalized by $g^2 m_\gamma^2/m_W^2$ 
for the primitive (red)
and improved (blue)
Ans\"atze.
Bounce actions obtained by the thin-wall approximation are shown as the dashed lines. The ratios of the mass parameters are indicated in the figures.
For comparison, we also show the bounce action in the infinitely thin string limit \eqref{eq:Preskill}, which becomes
\begin{align}
    S_B^{(\mathrm{thin})} \times g^2 \frac{m_\gamma^2}{m_W^2} = \frac{4\pi^2 f_M^2(\adjmass/m_W)}{f_T(m_{\fund_1}/m_\gamma)}\ .
\end{align}
The yellow band shows the 
range of the
bounce action compatible with the PTA data, namely
$S_B=\pi \kappa$
with $7.5<\sqrt{\kappa}<8.5$ 
for $g=1$ and $g=2$.

The figures indicate that
the asymptotic behavior of $S_B$ by the primitive Ansatz fairly reproduces that of
the infinitely thin approximation, $S_B^{(\mathrm{thin})}$.
In that region, we also found that the thin-wall approximation in \cref{eq:thinWALL} well approximates the 
full bounce solution, where 
$\meff$ matches with $m_M$ with an accuracy of several ten percents.
This is a success beyond what was expected in Ref.~\cite{Shifman:2002yi}.
As a result,  we confirmed that the infinitely thin limit approximation provides
a fair estimate of the bounce action
for $m_W \gg m_\gamma$.

The improved Ansatz, on the other hand,
results in much larger bounce action 
for $m_W \gg m_\gamma$.
The larger bounce action is due to the 
non-trivial kinetic function $\Keff$ which enhances the effective ``potential barrier.''
Thus, although the string tension
for given $\unw$ is smaller for the 
improved Ansatz (see \cref{fig:betatension}), the resultant 
bounce action is much larger than that of the primitive Ansatz.
However, this result may be due to our procedure of minimizing $\mathcal{K}_\mathrm{eff}(\unw)$
and $T(\unw)$ separately as explained above, rather than a problem inherent in the Ansatz itself.
Searches for a better procedure are left for a future work.

For parameter region with mild hierarchy,
i.e., $m_W/m_\gamma < O(10)$,
the thin-wall approximation deviates from the full bounce solution and 
underestimates the bounce action.
The deviation is due to the
rather low potential barrier 
compared to the difference of the ``vacuum energy,'' with which the thin wall approximation is no longer valid.
This observation signals that the infinitely thin approximation is also not valid for the mild hierarchy since
it corresponds to the thin wall approximation.

Besides, 
for the infinitely thin approximation, $S_B^{(\mathrm{thin})}$,
we use the monopole mass
and the string tension in the 
hierarchical limit, $m_W \gg m_\gamma$.
For $m_W/m_\gamma = O(1)$, 
on the other hand, 
the dynamics of $\phi$
and $\fund$ are not well separated, and hence, the expression of  $S_B^{(\mathrm{thin})}$ 
in \cref{eq:Preskill} is no more reliable.
Since the bounce action compatible with the PTA data requires $m_W/m_\gamma = O(1)$,
there is a large uncertainty
when interpreting bounce action in terms of monopole mass or string tension through $S_B^{(\mathrm{thin})}$.

The bounce actions obtained through the Ans\"atze, on the other hand, provide upper limits on the optimal bounce actions, since they connect the string configuration and the vacuum configurations.%
\footnote{Here, we assume that the string configuration with $\unw = 0$ is a stable and the lowest energy configuration even for $m_W/m_\gamma = O(1)$.}
Thus, when interpreting the 
bounce action compatible with the 
PTA data, our results 
provide reliable constraints 
on the model parameters even 
for $m_W/m_\gamma = O(1)$.%
\footnote{For $m_W/m_\gamma = O(1)$,  the mass parameters should not be taken as the physical masses of 
the particles but taken as the aliases of the coupling constants and the dimensionful parameters $\VG$
and $\VH$.}

Finally, we show the parameter dependence of the bounce action 
in \cref{fig:parameter-dependence}.
\Cref{fig:mass_dependence} shows the 
dependence on $m_{\fund_1}/m_\gamma$, $m_{\fund_2}/m_\gamma$ and 
$\adjmass/m_\gamma$.
The solid (dashed) lines show the dependence for the improved (primitive) Ansatz.
Here, we take $g=1$ and $m_W/m_\gamma = 5$.
The mass parameter is taken to be $\adjmass = m_W$, $m_{\fund_2} = m_W$ and $m_{\fund_1} = m_\gamma$, respectively, when not varied.
The bounce action becomes smaller for a larger $m_{\fund_1}/m_\gamma$,
while the dependencies on $m_{\fund_2}/m_\gamma$
and $\adjmass/m_\gamma$ are mild.

\Cref{fig:ratio_v_contour} shows the contour plot of the bounce action on $(V,m_{\fund_1}/m_\gamma)$ plane for $g=1$ and $v=1$.
The other mass parameters 
are set to $\adjmass=m_{\fund_2}=m_W$.
The solid lines are for the improved 
Ansatz and the dashed lines are 
for the bounce action by the Preskill-Vilenkin approximation.
The gray shaded region is 
classically unstable for the direction of the unwinding parameter $\unw$.
In the yellow shaded region, 
the bounce action in the improved Ansatz is smaller than $S_B^{(\mathrm{thin})}$, and hence,
$S_B^{(\mathrm{thin})}$
at least underestimates 
the string breaking rate.

As a guidance, we also show the 
parameter region 
which results in
$\sqrt{\kappa}\simeq \numrange{7.5}{8.5}$
in the improved Ansatz
as a cyan shaded region.
The light-cyan shaded region
corresponds to the same but with the Preskill-Vilenkin approximation.
Since the bounce action is the upper limit for given parameters,
the region upper left region
of the cyan shaded band cannot explain the PTA signal for $g=1$.

\section{Conclusions and Discussions}
In this paper, 
we revisited
the estimate of the string breaking rate
in a model with the two step symmetry breaking
$\SU(2)\to \U(1) \to \mbox{nothing}$ with
$\pi_2(\SU(2)/\U(1))=\pi_1(\U(1))=\mathbb{Z}$ while $\pi_2(\SU(2))=\pi_1(\SU(2))=0$.
Based on the analytical Ans\"atze
for the string unwinding process proposed in Ref.\,\cite{Shifman:2002yi}, 
we calculated the breaking rate 
which takes into account the finite sizes of the string and monopole.

As a result, we found that 
the asymptotic behavior of the 
bounce action for the primitive Ansatz well reproduces that of the 
bounce action in the infinitely thin approximation for $m_W \gg m_\gamma$.
This is a success beyond what was expected in Ref.~\cite{Shifman:2002yi},
which guarantees that the 
primitive Ansatz describes the 
breaking process via the (excited) monopole-antimonopole pair creation. 

Our analysis also revealed that 
the thin-wall approximation results in the underestimation of the bounce action for a mild hierarchy $m_W/m_\gamma = O(10)$ or below.
This deviation signals that the 
infinitely thin string approximation is also no more valid for $m_W/m_\gamma = O(10)$ or below. 

Note that for $m_W/m_\gamma = O(1)$, which is compatible with
the PTA data, the infinitely thin approximation is no longer reliable since the dynamics of $\phi$ and $\fund$ are not well separated.
Even in such a region, the bounce action obtained in this paper can be used 
to provide reliable constraints on the model parameters.

Finally, we enumerate points we left for future works.
Firstly, we did not show the classical stability of cosmic strings, especially for $m_W/m_\gamma = O(1)$.
At least we confirmed the classical stability in the direction of the unwinding parameter $\unw$ 
for $m_W/m_\gamma = O(1)$.
To show the classical stability,
however, we need to check
all the perturbations about the 
string configuration.
\modifiedat{JHEP2}{We also restricted ourselves to the symmetry breaking pattern $\SU(2)\to \U(1)\to 1$ and left the others to future work.
Notably, symmetry breaking chains of the type $\SU(2)\times \U(1)\to \U(1)\times \U(1)\to \U(1)$ often arise from e.g. SO(10) grand unified theories.
They differ from our case primarily in that the monopoles at the end of string segments carry unconfined U(1) flux.
Extension to such cases, which requires careful treatment of the unconfined magnetic field, will be given elsewhere.
}
We also did not discuss 
how the metastable string network is formed at the phase transition.
In particular, it is highly non-trivial
how many long strings are formed for $m_W/m_\gamma = O(1)$
where the phase transitions 
of $G\to$U(1) and U(1)$\to$nothing are not well separated.%
\footnote{As the thermal mass terms for the scalar fields are proportional to $\lambda$ and $\Vlambda$, we expect
that the symmetry breaking can be separated for $\lambda \gg \Vlambda$ even for $m_W/m_\gamma = O(1)$.}
To clarify this point, we need to perform cosmological lattice simulation, which we also leave for future work.

The bounce actions we obtained 
in this paper provide upper limits on the bounce action of the string breaking.
Thus, it is important to 
discuss whether there are
unwinding processes which could give a smaller bounce action.
For example, we find that 
linear combinations
of the primitive and the improved Ansatz can give 
an $O(10)$\% smaller bounce action.
We will update our analysis including the 
improvement of the minimization of 
the improved Ansatz in future work.

\section*{Acknowledgments}
This work is supported by Grant-in-Aid for Scientific Research from the Ministry of Education, Culture, Sports, Science, and Technology (MEXT), Japan, 21H04471, 22K03615 (M.I.), 20H01895, 20H05860 and 21H00067 (S.S.) and by World Premier International Research Center Initiative (WPI), MEXT, Japan. 
This work is also supported by Grant-in-Aid for JSPS Research Fellow 
JP22KJ0556 (Y.N.) 
and by International Graduate Program for Excellence in Earth-Space Science (Y.N.).
This work is supported by JST SPRING Grant Number JPMJSP2108 (K.W.).
This work is supported by FoPM, WINGS Program, the University of Tokyo (A.C.).

\appendix

\section{Long Expressions}
\label{sec:longeqs}
For completeness, here we list long expressions that are too complicated and unilluminating to be included in the body.

For the primitive Ansatz,
\begin{align}\label{eq:Hprimitive}
    \hamiltonian_\unw = 8\pi \rho\Biggl[{}&
    \frac{1}{g^2}(\delrho a_\unw)^2 \nonumber\\
    &\begin{aligned}
    -a_\unw\biggl\{&-2\adjprofile_\unw(V\sin 2\unw - \adjprofile_\unw)+\frac{1}{2}V (\partial_\unw \adjprofile_\unw)\cos 2\unw\\
    &+ \frac{1}{g^2\rho^2}\qty(f_\unw^2-\frac{1}{2}(\partial_\unw f_\unw)\sin 2\unw  - f_\unw (1-f_\unw)\cos 2\unw )\biggr\}
    \end{aligned} \nonumber\\
    &+a_\unw^2\Bqty{\frac{1}{g^2\rho^2}(1-4f_\unw(1-f_\unw)\cos^2\unw
    )+V^2+\frac{1}{2}\higgsprofile_\unw^2+\adjprofile_\unw (-2V\sin 2\unw + \adjprofile_\unw)}
    \Biggr]\ ,
\end{align}
\begin{multline}\label{eq:primitiveK}
K(\unw,\rho)=\hamiltonian_\unw + \frac{2\pi}{g^2 \rho} \Bigl[2 (\delunw f_{\unw })^2-4  f_{\unw } \delunw f_{\unw }\sin 2 \unw+2 (\delunw f_{\unw })^2\cos 2 \unw \\+2 g^2 \rho^2( \delunw \higgsprofile_{\unw })^2+g^2 \rho^2 (\delunw \adjprofile _{\unw })^2+4 f_{\unw }{}^2+4 g^2 \rho^2 \adjprofile _{\unw }{}^2\Bigr]\ ,
\end{multline}
and
\begin{multline}\label{eq:primitiveV}
    V(\unw,\rho)=\frac{1}{2} \pi  \rho \Biggl(\frac{8 \cos ^2\unw (\delrho f_{\unw })^2}{g^2 \rho^2}+4 (\delrho \higgsprofile_{\unw })^2+2 (\delrho \adjprofile_{\unw })^2+\frac{4 \cos ^2\unw f_{\unw }{}^2 \higgsprofile_{\unw }{}^2}{\rho^2}\\
    +\frac{8 \cos ^2\unw f_{\unw }{}^2 \left(V \sin \unw-\cos \unw \adjprofile _{\unw }\right){}^2}{\rho^2}+\frac{4 (\cos \unw+\cos 3 \unw ) f_{\unw } \adjprofile _{\unw } \left(V \sin \unw-\cos \unw \adjprofile _{\unw }\right)}{\rho^2}\\
    +4 \Hlambda  \higgsprofile_{\unw }{}^4-\higgsprofile_{\unw }{}^2 \left(8 \Hlambda  \VH^2-\gamma  \adjprofile _{\unw }{}^2\right)+\frac{(\cos 4 \unw +1) \adjprofile _{\unw }{}^2}{\rho^2}+4 \Vlambda  \adjprofile _{\unw }{}^2 \left(\adjprofile _{\unw }-2 \VG \sin 2 \unw \right){}^2+4 \lambda  \VH^4\Biggr)\ .
\end{multline}

For the improved ansatz,
\begin{align}
    \hamiltonian_\unw=\frac{4 \pi}{g^2 \rho}\Bigl[&2 \left(
    \partial_{\rho}
    a_{\unw}
\right){}^2 \rho^2\nonumber\\
    &\begin{aligned}
    +a_{\unw}{}^2\Bigl\{&8 f^{\gamma}_{\unw} \cos ^2\unw \left(f^{\gamma}_{\unw} \cos ^2\unw-1\right)+2 \left((f^W_{\unw})^2 \sin ^22 \unw +g^2 \VG^2 \rho^2+1\right)\\
    &+g^2 \rho^2 \left(\higgsprofile_{\unw}{}^2+2 \Phi_{\unw} (\Phi_{\unw}-2 \VG \sin 2 \unw)\right)\Bigr\}
    \end{aligned}\nonumber\\
    &\begin{aligned}
    +2 a_{\unw} \Bigl\{&2 f^W_{\unw} \left(\cos ^2\unw \left(\left(
    \partial_{\unw}
    f^{\gamma}_{\unw}
\right) \sin 2 \unw-2 f^{\gamma}_{\unw}\right)+\cos 2 \unw\right)\\
&+\left(
    \partial_{\unw}
    f^W_{\unw}
\right) \left(\sin 2 \unw-4 f^{\gamma}_{\unw} \sin \unw \cos ^3\unw\right)\\
&-g^2 \VG \left(
    \partial_{\unw}
    \Phi_{\unw}
\right) \rho^2 \cos 2 \unw+2 g^2 \rho^2 \Phi_{\unw} (V \sin 2 \unw-\Phi_{\unw})\Bigr\} \Bigr]\ ,
\end{aligned}
\end{align}
\begin{multline}
\label{eq: improveK}
   K(\unw,\rho)=\hamiltonian_\unw+\frac{2\pi}{g^2 \rho}\Bigl[4 \cos ^2\unw \qty(\left(
    \partial_{\unw}
    f^{\gamma}_{\unw}
\right) \cos \unw-2 f^{\gamma}_{\unw} \sin \unw){}^2\\
+(2 f^W_{\unw} \cos 2 \unw+\left(
    \partial_{\unw}
    f^W_{\unw}
\right) \sin 2 \unw){}^2
+g^2 \rho^2 \left(2 \left(
    \partial_{\unw}
    \higgsprofile_{\unw}
\right){}^2+\left(
    \partial_{\unw}
    \Phi_{\unw}
\right){}^2+4 \Phi_{\unw}{}^2\right)\Bigr]\ ,
\end{multline}
and 
\begin{multline}
\label{eq: improveV}
    V(\unw,\rho)=\frac{\pi}{2 \rho} \Bigl[\higgsprofile_{\unw}{}^2 \left(4 \pqty{f^{\gamma}_{\unw}}^2 \cos ^4\unw+\pqty{f^W_{\unw}}^2 \sin ^22 \unw+\rho^2 \left(\gamma  \Phi_{\unw}{}^2-8 \lambda  \VH^2\right)\right)\\
    +2 f^W_{\unw} \Phi_{\unw} \sin 4 \unw \left(2 f^{\gamma}_{\unw} \cos ^2\unw-1\right) (\Phi_{\unw} \sin 2 \unw-V)\\
    +\Phi_{\unw}{}^2 \left(2 f^{\gamma}_{\unw} (\cos \unw+\cos 3 \unw)^2 \left(f^{\gamma}_{\unw} \cos ^2\unw-1\right)+\cos 4 \unw \left(1-8 \VG^2 \rho^2 \Vlambda\right)+8 \VG^2 \rho^2 \Vlambda+1\right)\\
    +2 \pqty{f^W_{\unw}}^2 \sin ^22 \unw (V-\Phi_{\unw} \sin 2 \unw){}^2\\
    +\frac{1}{g^2}\pqty{2 \left(g^2 \rho^2 \left(2 \left(
    \partial_{\rho}
    \higgsprofile_{\unw}
\right){}^2+\left(
    \partial_{\rho}
    \Phi_{\unw}
\right){}^2+2 \lambda  v^4\right)+4 \left(
    \partial_{\rho}
    f^{\gamma}_{\unw}
\right){}^2 \cos ^4\unw+\left(
    \partial_{\rho}
    f^W_{\unw}
\right){}^2 \sin ^22 \unw\right)}\\
+4 \lambda  \higgsprofile_{\unw}{}^4 \rho^2-16 \VG \rho^2 \Vlambda \Phi_{\unw}{}^3 \sin 2 \unw+4 \rho^2 \Vlambda \Phi_{\unw}{}^4\Bigr]\ .
\end{multline}

\bibliographystyle{apsrev4-1}
\bibliography{bibtex}

\end{document}